\documentclass[preprint,showpacs,preprintnumbers,amsmath,amssymb,superscriptaddress]{revtex4}
\usepackage{graphicx} 
\usepackage{dcolumn}  
\usepackage{bm}       

\begin{document}

\title{Even-odd effects and Coulomb effects on minimal excitation energy of fragments from low energy fission}
\author{M. Montoya}
\email{mmontoya@ipen.gob.pe}

\affiliation{Instituto Peruano de Energ\'ia Nuclear, Canad\'a 1470, San Borja, Lima, Per\'u.}
\affiliation{Facultad de Ciencias, Universidad Nacional de Ingenier\'ia, Av. T\'upac Amaru 210, R\'imac, Lima, Per\'u.}
\begin{abstract}
This work is focused on even-odd effects on the minimal total fragment excitation energy in thermal neutron induced fission of $^{233}$U and $^{235}$U as well as in spontaneous fission of $^{252}$Cf. In a scission model, taking into account the fragment deformation properties and Coulomb interaction between fragments, the expression of the difference between $Q$-values referred to even/even and odd/odd charge splits, respectively,  on the corresponding difference between the minimal total fragment excitation energy  is studied.
\end{abstract}

\maketitle

\section{Introduction}
In 1969, A. C. Wahl {\it et al.} reported the observation of a preference for even/even charge splits from $^{239}$Pu(n$_{\rm{th}}$, f) \cite{Wahl1969}. This so called even-odd effect on charge distribution was interpreted as an indicator of an intrinsic excitation of the system during the descent from saddle to scission. In 1980, W. Lang {\it et al.} reported that the mentioned even-odd effect increases with kinetic energy of fragments from thermal neutron induced fission of $^{235}$U \cite{Lang1980}. 
Contrary to expected, in 1981, C. Signarbieux {\it et al.} obtained experimental evidence that in the region of highest kinetic energy of fragments from  $^{233}$U(n$_{\rm{th}}$, f) and $^{235}$U(n$_{\rm{th}}$, f), respectively, there is no evidence of even-odd effects on mass yield \cite{Signarbieux1981}.  Similar results were found in 1992 by H.-H. Knitter {\it et al.} on mass and charge distribution of fragments from $^{252}$Cf(sf) \cite{Knitter1992}. 

In 1993, F.-J. Hambsch {\it et al.} present the charge and mass distribution as a function of excitation energy of fragments from  $^{252}$Cf(sf) \cite{Hambsch1993}. They reported the observation of a preference for odd over  even charge fragments in the region of low values of total excitation energy. The question that emerges from these results is the following: what happens in the case of minimal value of total excitation energy of fragments? Do the odd charge fragments reach lower minimal total excitation energy than the even charge fragments do? If that is the case, what is the physical cause?
\section{Even-odd effects on minimal total excitation energy}

\begin{figure}
\centering
\includegraphics[width=12cm]{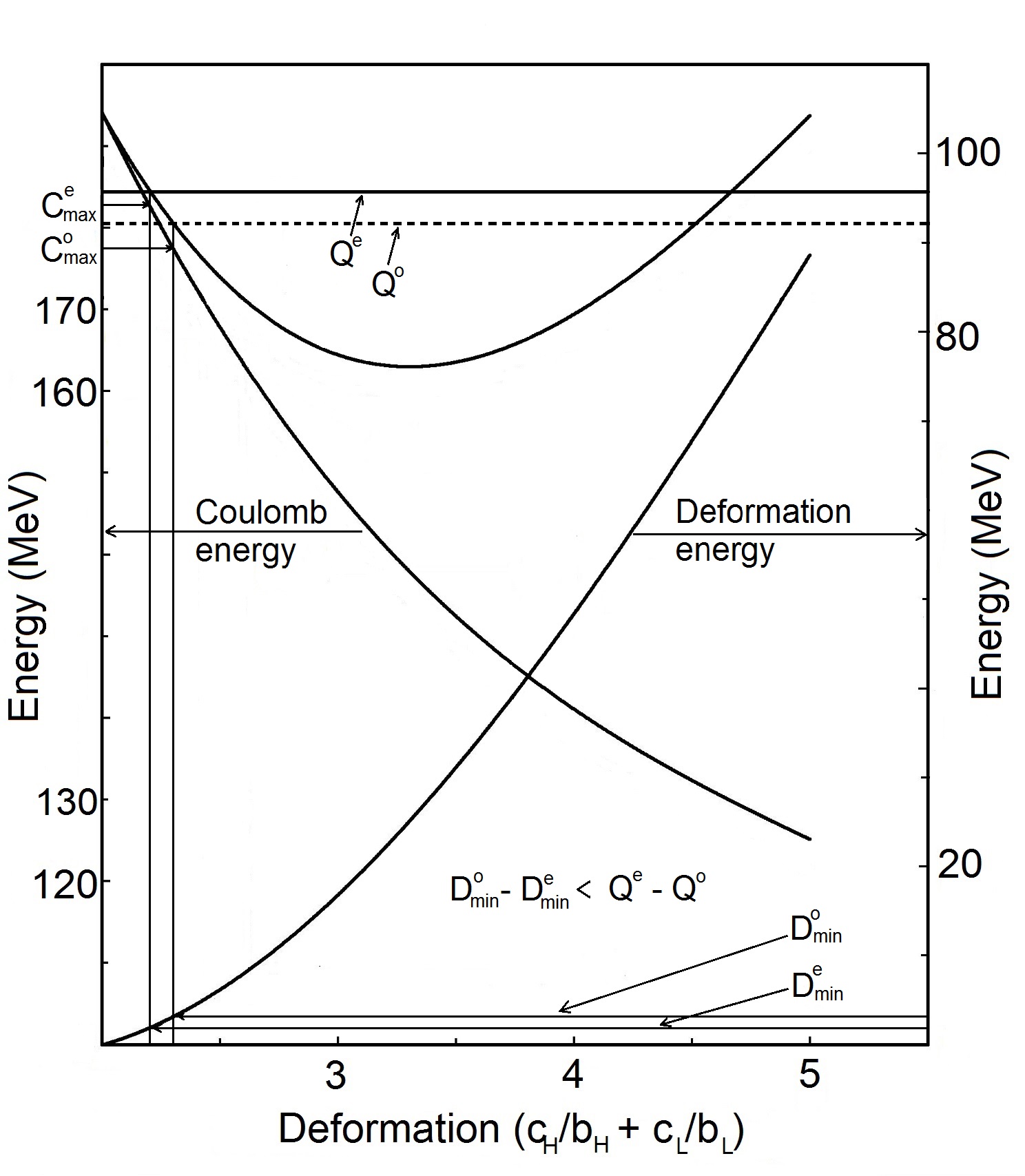}
\caption{Low energy fission of actinides. Schematically, solid lines represent the total deformation energy ($D$), the Coulomb interaction energy ($C$) and the potential energy  ($P=D+C$) as a function of the elongation of a scission configuration.  The space of deformation is limited by the total available energy ($Q$). Horizontal solid line and dashed lines represent the $Q$-value corresponding to even and charge splits, respectively.}
\label{fig:fig1.jpg}
\end{figure}

In the scission configuration, the equation of energy balance is the following:
\begin{equation}
Q = P + X^*  + K_{\rm{free}},
\end{equation}
where $Q$ is the available energy, $P$ is the potential energy, $X^*$ is the total intrinsic excitation energy, $K_{\rm{free}}$ is the total free pre-scission kinetic energy. The potential energy is given by
\begin{equation}
P\left({\mathcal D}\right) = C\left({\mathcal D}\right) + D\left({\mathcal D}\right),
\end{equation}
where $C$ is the energy of Coulomb interaction energy between fragments,  
$D$ is the total deformation energy and ${\mathcal D}$ represents the elongation of the scission configuration, which may be defined as:
\begin{equation}
\mathcal{D} =  \frac{c_{\rm{H}}}{b_{\rm{H}}} + \frac{c_{\rm{L}}}{b_{\rm{L}}},
\end{equation}
where $b$ and $c$ are the ellipsoidal semi axis referred to heavy (H) and light (L) fragments, respectively. It is assumed that for both light and heavy fragments $a$ = $b$. See Fig. ~\ref{fig:fig1.jpg}.
The aim of this work is to study how the components of that equation influence the even-odd effects on the minimal excitation energy. Let's assume that the minimal excitation energy is reached by the initially most compact scission configuration with null pre-scission kinetic energy and null intrinsic excitation energy. Then the minimal total deformation energy which will correspond to the minimal excitation deformation,  obeys the relation
\begin{equation}
D_{\rm{min}}=Q-C_{\rm{max}},
\end{equation}
where $C_{\rm{max}}$ is the maximal value of Coulomb interaction energy and $D_{\rm{min}}$ is the minimal value of deformation energy reached by the scission configuration.\\
Let´s define:
\begin{equation}
\Delta Q = Q^{\rm{e}}- Q^{\rm{o}},
\end{equation}
\begin{equation}
\Delta D_{\rm{min}}= D_{\rm{min}}^{\rm{e}} - D_{\rm{min}}^{\rm{o}},
\end{equation}
and
\begin{equation}
ΔC_{\rm{max}}= C_{\rm{max}}^{\rm{e}} - C_{\rm{max}}^{\rm{o}}.
\end{equation}
Therefore the difference between total deformation energies corresponding to even/even and odd/odd charge splits, respectively, will be:
\begin{equation}
\Delta D_{\rm{min}}= \Delta Q - \Delta C_{\rm{max}}.
\end{equation}
\\
Let's assume that $D$ increases with elongation of the scission configuration. Then 
\begin{equation}
\Delta D_{\rm{min}}  < 0.
\end{equation}
The amplitude of $\Delta D_{\rm{min}}$ will increase with the configuration hardness. In other words the tendency will be in the sense that even/even charge splits will have lower minimal deformation energy than the corresponding to odd/odd charge splits.\\
In cases of soft scission cold configurations, i.e. the deformation energy does not change with deformation in cold fission region,  there will be not even-odd effects on $D_{\rm{min}}$, i.e.
\begin{equation}
\Delta D_{\rm{min}} = 0.
\end{equation}
In this case the even-odd effect on $Q$ value will be expressed on $C_{\rm{max}}$, i.e.
\begin{equation}
\Delta C_{\rm{max}} =  \Delta Q,
\end{equation}
therefore on $K_{\rm{max}}$.
\section{Coulomb effects on minimal total excitation energy}
\begin{figure}
\centering
\includegraphics[width=12cm]{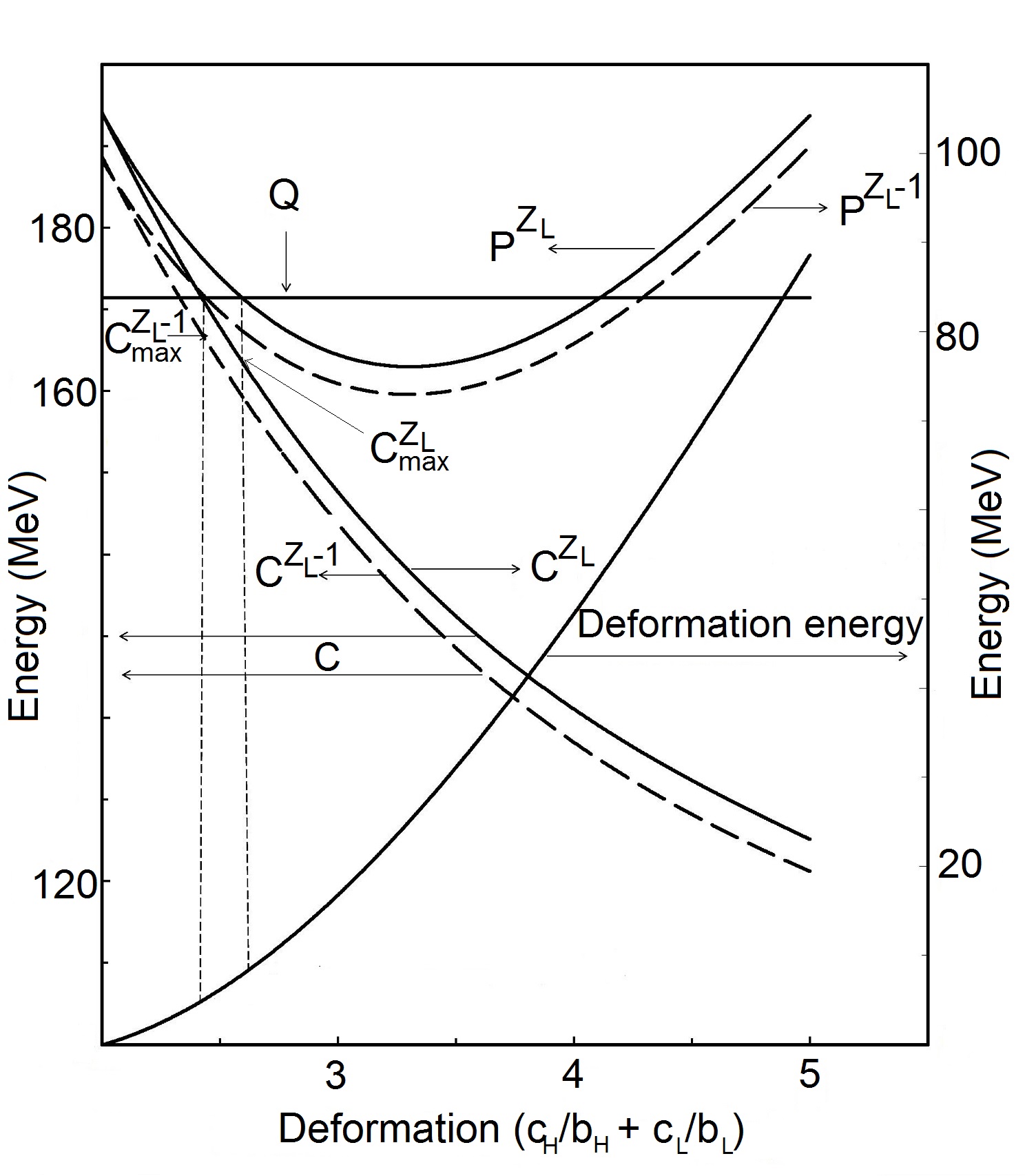}
\caption{Low energy fission of actinides. Schematically, solid lines represent the total deformation energy ($D$), the Coulomb interaction energy ($C$) and the potential energy  ($P=D+C$) as a function of the elongation of a scission configuration, corresponding to light fragment charge $Z_{\rm{L}}$. The space of deformation is limited by the total available energy ($Q$). Dashed lines represent similar curves corresponding to the neighbouring more asymmetrical charge split ($Z_{\rm{L}} - 1$) but having the same $Q$-value. One can see that the lower minimal deformation corresponds to the lower light fragment charge. As a result, the higher maximal Coulomb interaction energy, which will be converted in kinetic energy, corresponds to charge $Z_{\rm{L}}-1$.}
\label{fig:fig2.jpg}
\end{figure}

The previous section was dedicated to analyse even-odd effect considering that for the same deformation Coulomb energy does not change with fragment charge. However, when two neighboring fragment charge are compared, the Coulomb effect must be taken into account \cite{Montoya1984, Montoya1986, Montoya2014}. In a scission point model, the potential energy ($P$) of a scission configuration corresponding to a light fragment charge $Z_{\rm{L}}$ is given by the relation
\begin{equation}
P^{Z_{\rm{L}}}\left({\mathcal D}\right)=D^{Z_{\rm{L}}}\left({\mathcal D}\right)+C^{Z_{\rm{L}}}\left({\mathcal D}\right).
\end{equation}
See Fig. ~\ref{fig:fig2.jpg}.\\
One assumes that the most compact configuration obeys the relations $E_{\rm free}=0$ and $X^*=0$. Then
\begin{equation}
Q = P.
\end{equation}

After the Coulomb effect it is expected that between two isobaric splits having similar $Q$-values, the more asymmetric charge  split will reach a more compact configuration, which corresponds to a lower deformation energy \cite{Montoya2014}. Because Coulomb effect may involve more than 3 MeV, for experimental results that can not be interpreted with Coulomb effect hypothesis an eventual dramatic change of deformation properties for neighboring charge isobaric fragments may be explored.\\ 
In cases of soft scission configurations, i.e. the deformation energy does not vary with deformation,  there will be not Coulomb effect on $D_{\rm{min}}$, i.e.
\begin{equation}
D^{Z_{\rm{L}}-1}_{\rm{min}}=D^{Z_{\rm{L}}}_{\rm{min}}.
\end{equation}
\section{Analysis of experimental results about minimal excitation energy in isobaric charge splits}
\begin{figure}
\centering
\includegraphics[width=12cm]{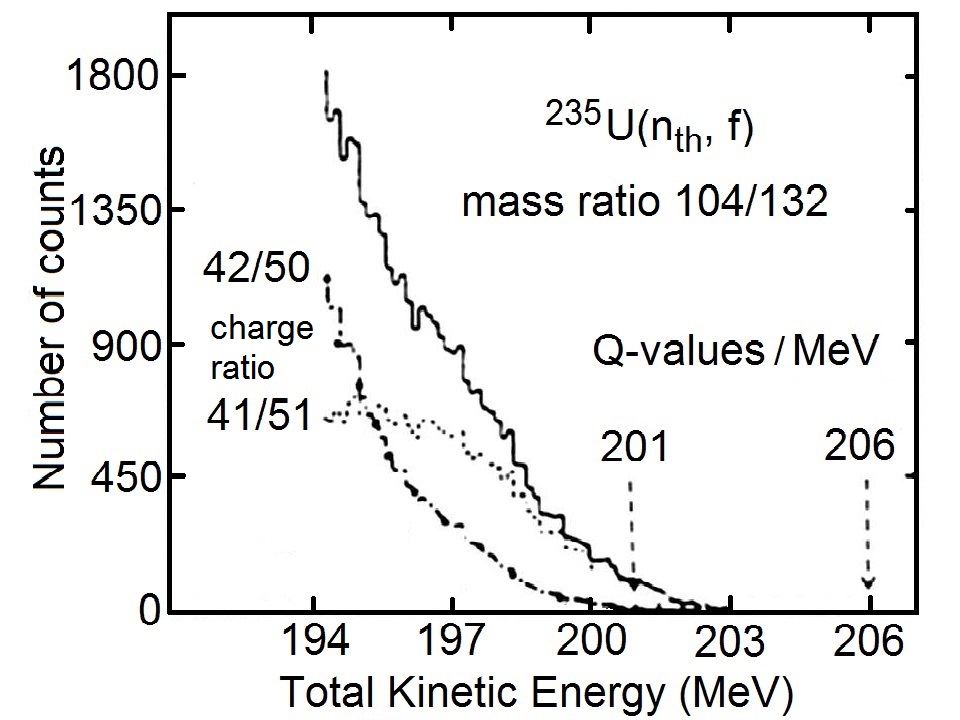}
\caption{High K tail for the mass split 104/132 resolved into the charge splits 42/50 and 41/51, respectively, from $^{235}$U(n$_{\rm{th}}$, f). Taken from Ref.~\cite{Gonnenwein2013}.}
\label{fig:fig3.jpg}
\end{figure}

Based on experimental results obtained in 1989 by J. Trochon {\it et al.} \cite{Trochon1989}, in 2013 F. G{\"o}nnenwein presents the yield of charge as a function of total kinetic energy ($K$) of isobaric fragmentations $A_{L}/A_{H} = 104/132$ from $^{235}$U(n$_{\rm{th}}$, f). See Fig. ~\ref{fig:fig3.jpg}. He observes that the minimal total excitation energy is null for odd/odd charge splits 41/51; while the corresponding to the even/even charge split 42/50 is higher than 3 MeV \cite{Gonnenwein2013}. F. G{\"o}nnenwein interprets those results as a consequence of the higher density of low energy levels in odd charge fragments. However the above mentioned experimental results are also compatible with the hypothesis of Coulomb effect: the more asymmetric charge split reaches the lower minimal deformation energy.\\
Based on experimental results about $^{252}$Cf(sf) reported by Cr{\"o}ni {\it et al.} in 1997 \cite{Croni1997}, referred to isobaric fragmentation $A_{L}/A_{H} = 120/132$, F. G{\"o}nnenwein notices that $K$ for odd/odd charge split 47/51 reaches the corresponding $Q$-value, while the $K$ values referred to the even/even charge split = 48/50 stay away from its $Q$-value \cite{Gonnenwein2012}. See Fig. ~\ref{fig:fig5.jpg}. Similarly to the previous case, one can say that in this case also the more asymmetric charge split reaches the lower minimal deformation energy.\\
For the isobaric fragmentations $A_{L}/A_{H} = 118/134$, F. G{\"o}nnenwein notices that $K_{\rm{max}}$ corresponding to the odd/odd charge split 47/51 is about 4 MeV lower than the corresponding $Q$-value, while $K_{\rm{max}}$  referred to the more asymmetric even/even charge split 46/52 is only about 2 MeV lower than the $Q$-value. See Fig. ~\ref{fig:fig5.jpg}. This result is also compatible with the Coulomb effect hypothesis: the more asymmetric charge split reaches the lower minimal deformation energy. \\
In 1994 W. Schwab {\it et al.} presented results referred to $^{233}$U(n$_{\rm{th}}$, f) \cite{Schwab1994}. For the isobaric fragmentation $A_{L}/A_{H} = 86/138$, the minimal excitation energy reached by the charge split 34/58 is lower than the corresponding to the charge split 35/57 and this is lower than the corresponding to the charge split 36/56.  See Fig. ~\ref{fig:fig6.jpg}. Again this result is coherent with Coulomb effect hypothesis.\\
For the isobaric fragmentation $A_{L}/A_{H} = 88/146$ from  $^{233}$U(n$_{\rm{th}}$, f) the even/even charge split 36/56 reaches lower excitation energy than the odd/odd charge split 37/55. See Fig. ~\ref{fig:fig7.jpg}. In this case the even/even and more asymmetric charge split reaches lower excitation energy, result which also agrees with Coulomb effect hypothesis. \\
In a global view, even-odd effects on measured fragment charge referred to maximal value of total kinetic energy are observed in isobaric fragmentations from $^{235}$U(n$_{\rm{th}}$,f) reported by C. Signarbieux \cite{Signarbieux1991}. Between the 29 cases of isobaric fragmentations, in 25 cases the highest values of total kinetic energy correspond to even/even charge splits. Nevertheless, when two isobaric charge splits with similar $Q$-value are compared, the more asymmetric split reaches their corresponding maximal $K$-values. 

\section{Conclusion}
Several experimental results presented by other authors seem to show that odd/odd charge splits reach lower minimal excitation energy than the even/even charge splits do. However the presented cases correspond to odd/odd charge splits more asymmetric than the  even/even charges splits. Therefore those results may be interpreted by the Coulomb effect hypothesis. After this hypothesis the more asymmetric charge split will reach the lower minimal deformation energy. Moreover, there are case of even/even charge splits with higher minimal excitation energy than the odd/odd charge splits, but they also correspond to more asymmetric charge splits. Is short Coulomb effect seems to overshadow even-odd effects on minimal value of excitation energy.

\begin{figure}
\centering
\includegraphics[width=12cm]{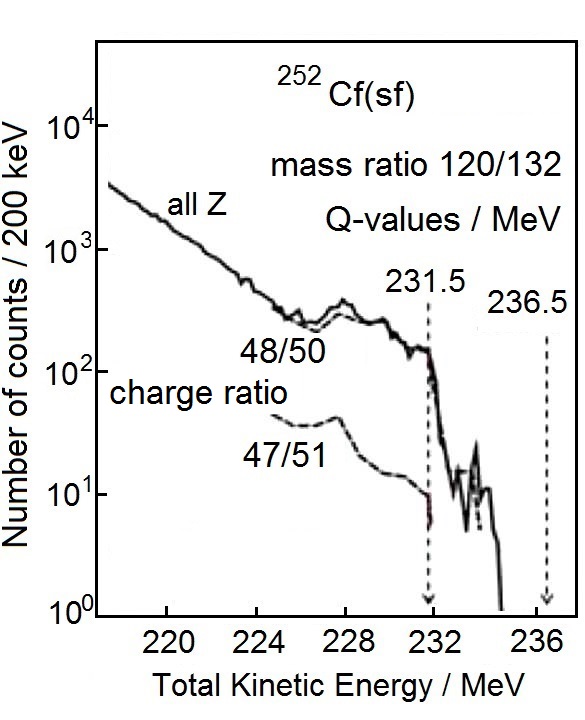}
\caption{High K tails for the mass ratio 120/132 in $^{252}$Cf(sf). The contributions of charge splits 48/50 and 47/51, respectively, and their corresponding $Q$-values are shown. Taken from Ref.~\cite{Gonnenwein2012}.}
\label{fig:fig4.jpg}
\end{figure}
\begin{figure}
\centering
\includegraphics[width=12cm]{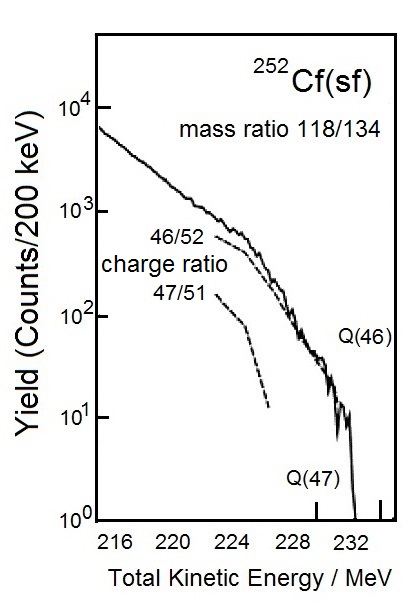}
\caption{High K tails for the mass ratio 118/134 in $^{252}$Cf(sf). The contributions of charge splits 46/52 and 47/51, respectively, and their corresponding $Q$-values are shown.Taken from Ref.~\cite{Gonnenwein2012}.}
\label{fig:fig5.jpg}
\end{figure}
\begin{figure}
\centering
\includegraphics[width=8cm]{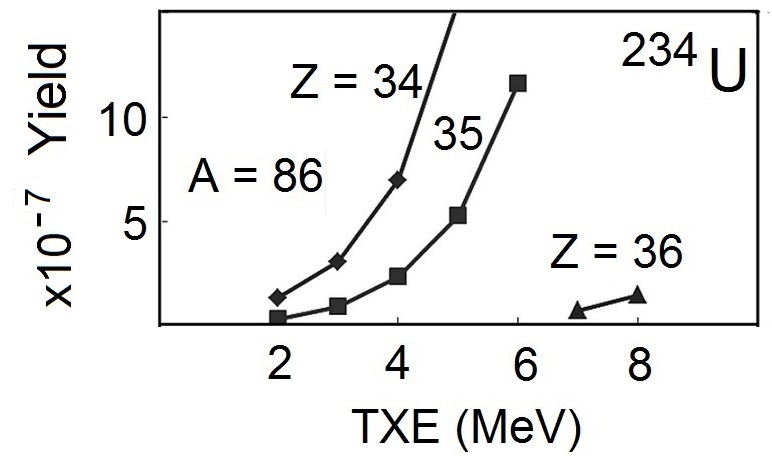}
\caption{Yield of charge as a function of excitation energy for  light fragment mass 86 from $^{233}$U(n$_{\rm{th}}$, f). Data taken from \cite{Schwab1994}.}
\label{fig:fig6.jpg}
\end{figure}
\begin{figure}
\centering
\includegraphics[width=8cm]{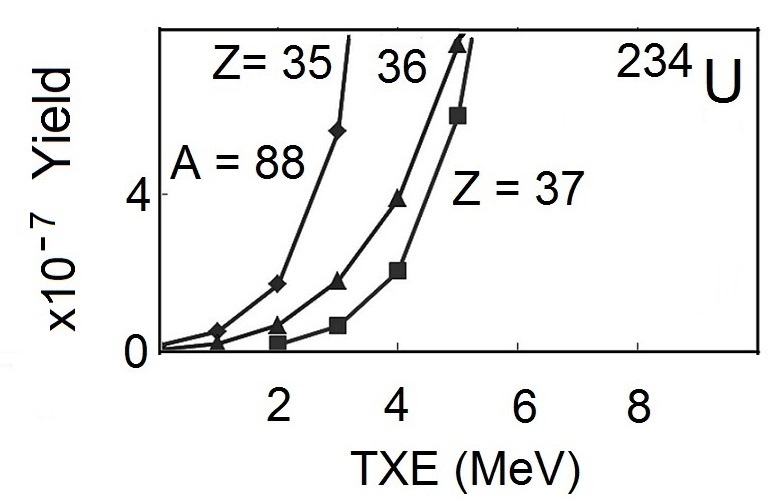}
\caption{Yield of charge as a function of excitation energy for  light fragment mass 88 from $^{233}$U(n$_{\rm{th}}$, f). Data taken from \cite{Schwab1994}.}
\label{fig:fig7.jpg}
\end{figure}

\section*{References}
\begin{thebibliography}{mylit}
\bibitem{Wahl1969} A. C. Wahl, A. E. Norris, R. A. Rouse and J. C. Williams, Proc. Symp. On Physics and Chemistry of Fission, (AEIA, Vienna, 1969) p. 813.
\bibitem{Lang1980} W. Lang, H.G. Clerc, H. Wohlfarth, H. Schrader, K.H Schmidt, Nucl. Phys. A345 (1980) 34
\bibitem{Signarbieux1981} C. Signarbieux, M. Montoya, M. Ribrag, C. Mazur, C. Guet, P. Perrin and M. Maurel, J. de Phys.Lett. 42 (1981) L437
\bibitem{Knitter1992} H.-H. Knitter, F.-J. Hambsch and C. Budtz-Jorgensen, Nucl. Phys. A536 (1992) 221
\bibitem{Hambsch1993} F.-J. Hambsch, H.-H. Knitter and C. Budtz-Jorgensen, Nucl. Phys. A554 (1993) 209-222
\bibitem{Montoya1984} M. Montoya, Z. Phys. A – Atoms and Nuclei 319 (1984) 219-225
\bibitem{Montoya1986} M. Montoya, R.W. Hasse and P. Koczon, Z. Phys. A – Atoms and Nuclei 325 (1986) 357-362
\bibitem{Montoya2014} M. Montoya, Rev. Mex. Fis. 60(5) (2014) 350.
\bibitem{Trochon1989} J. Trochon, G. Simon and C. Signarbieux, “50 Years with Nuclear Fission”, Am. Nucl. Soc., 1989, 313
\bibitem{Gonnenwein2013} F. G{\"o}nnenwein, Physics Procedia 47 ( 2013 ) 107-114

\bibitem{Croni1997} M. Cr{\"o}ni, A. M{\"o}ller, A. K{\"o}tzle, F. G{\"o}nnenwein, A. Gagarski and G. Petrov, “Fission and properties of neutron-rich nuclei”, World Scientific, 1997, 109
\bibitem{Gonnenwein2012} F. G{\"o}nnenwein in “The Neutron” by H. B{\"o}rner and F. G{\"o}nnenwein, World Scientific 2012
\bibitem{Schwab1994} W. Schwab, H.-G. Clerc, M. Mutterer, J.E Theobald, H. Faust, Nucl. Phys. A577 (1994) 674-690
\bibitem{Signarbieux1991} C. Signarbieux, Cold fragmentation properties: a crucial test for dynamics of nuclear fission, Proceedings of a Intern. Workshop. on Dynamical Aspects
of Nuclear Fission, Smolenice, J.I.N.R., Dubna  (1991), p. 19

\end{thebibliography}
\end{document}